\documentstyle[preprint,aps,tighten,epsfig]{revtex}

\def\q1{{q^{-1}}}
\def\qq1{{q-q^{-1}}}
\def\djk{{\partial_x^{(q)}}}
\def\dqq{{{\cal D}_x^{(q)}}}

\begin{document}
\draft
\title{Generalized symmetric nonextensive thermostatistics and
$q$-modified structures}
\author{A. Lavagno $^a$ and P. Narayana Swamy $^b$}
\address{
$^a$ Dipartimento di Fisica and INFN, Politecnico di Torino, 
I-10129 Torino, Italy\\
$^b$ Physics  Department, Southern Illinois University,
Edwardsville, IL 62026, USA}
\maketitle

\begin{abstract}
We formulate a convenient generalization of
the $q$-expectation value, based on the analogy of the symmetric
quantum groups and $q$-calculus, and  show that 
the  $q\leftrightarrow \q1$ symmetric
nonextensive entropy preserves all of the  mathematical structure of
thermodynamics  just as in the case of non-symmetric Tsallis
statistics. Basic properties and analogies with
quantum groups are discussed.
\end{abstract}

\vspace{0.5cm}

\vspace{1cm}

In the last few years there has been much interest in
nonextensive classical and quantum  physics.
The nonextensive statistical mechanics proposed by Tsallis
\cite{tsallis,curado},
has been
the source of inspiration for many investigations  in systems
 which represent multifractal properties, long-range interactions
and/or long-range memory effects \cite{biblio}. On the other hand,
 quantum groups and the derived $q$-deformed algebraic structure such as
$q$-oscillators, based on
the deformation of standard oscillator commutation-anticommutation relations, 
 have created considerable interest in mathematical physics
and in several applications \cite{bie}.
The investigations described above are 
only two apparently unrelated areas in nonextensive physics.
Although a complete understanding of the connection between nonextensive
statistics and $q$-deformed structure is still lacking, many papers are
devoted to the study of a deep connection between these two non-extensive
formalisms \cite{tsa,abe,bor,pin,joh}.

Tsallis statistics is a $q$-nonsymmetric formalism i.e., not invariant
under $q \leftrightarrow\q1$. Recently Abe \cite{abe} has
employed a connection between Tsallis entropy and the non-symmetric Jackson
derivative. Because the requirement of invariance under
$q \leftrightarrow\q1$ is very important in quantum groups \cite{aiz}, 
the above connection allows him to extend the  Tsallis entropy
to the $q$-symmetric one by means of a symmetric Jackson derivative.
However, Abe has not extended the definition of the expectation value
of an observable to the symmetric case and thus unable to formulate the
thermostatistics which will preserve the Legendre transformation of
standard thermodynamics 
in contrast to the  Tsallis statistics which does. We would like to point
out that Ref.\cite{bor} introduces a two-parameter modification for the
entropy and for the expectation value of an observable but does not also
produce a consistent formulation of thermostatistics.  

The purpose of this letter is to show how the $q\leftrightarrow \q1$
symmetric generalization of the Tsallis entropy
together with a natural generalization of the
$q$-expectation value produces  
 a thermostatistics that preserves the
mathematical structure of standard thermodynamics and show
that this
property is a direct consequence of the generalization from the
non-symmetric $q$-calculus to the symmetric one.

Before  investigating the symmetric nonextensive thermostatistics, let us
briefly review the fundamental properties of the Tsallis thermostatistics,
which is based  upon the following two postulates \cite{tsallis,curado}.

\begin{itemize}
\item
A nonextensive generalization of the Boltzmann-Gibbs entropy
(Boltzmann constant is set equal to unity)
\begin{equation}\label{s}
S_q =\frac{1}{q-1}  \left ( 1-\sum_{i=1}^W p_i^q\right ) \; ,
\ \ \ {\rm with} \ \ \  \sum_{i=1}^W p_i=1  \; ,
\end{equation}
where $p_i$ is the probability of a given microstate among $W$ different ones
and $q$ is a fixed real parameter.
The new entropy has the usual properties of positivity, equiprobability,
and reduces to the conventional Boltzmann--Gibbs entropy
$S=-\sum_i p_i \ln p_i$ in the limit $q\rightarrow 1$.

\item
A generalized definition of internal energy
\begin{equation}
 U_q  = \sum_i \epsilon_i \; p_i^q
\label{uq}
\end{equation}
and, accordingly, a generalization of the $q$-expectation value of
an observable $A$ which can be expressed as  
$\langle A \rangle_q \equiv \sum_i A_i \; p_i^q$.
In the limit $q\rightarrow 1$,  
$\langle A \rangle_1$ corresponds to the standard mean value.
This postulate plays a central role in the derivation
of the equilibrium distribution and leads to the correct thermodynamic relations.

\end{itemize}

The deformation parameter $q$ measures the degree of nonextensivity of the
theory. In fact, if we have two independent systems $A$ e $B$, such that
the probability of $A+B$ is factorized into
$p_{A+B}(u_A, u_B)=p_A(u_A) \, p_B(u_B)$, the global entropy is not simply
the sum of  their entropies and it is easy to verify that
\begin{equation}
S_q(A+B)=S_q(A)+S_q(B)+(1-q) S_q(A)S_q(B) \; .
\label{add}
\end{equation}

Another important property is that $S_q$ is consistent with Laplace's
maximum ignorance principle, i.e., if 
$p_i= 1/W, \;\;\forall{i}$ and $W \ge 1$ (equiprobability) one has the
following extremum value \cite{curado}
\begin{equation}
S_q= \ln_q W \; ,
\label{boltz}
\end{equation}
where we have defined the generalized logarithmic function $\ln_q x$ as
\begin{equation}
\ln_q x=\frac{x^{1-q}-1}{1-q} \;  .
\label{logq}
\end{equation}
In the limit $q \rightarrow 1$, $\ln_q x\rightarrow \ln x$ and
Eq.(\ref{boltz}) reproduces Boltzmann's
celebrated formula $S=\ln\; W$. Let us stress the
importance of this result for the purpose of the following
discussion, because it defines the generalized logarithmic
function in nonextensive statistics and plays a crucial role in the
determination of the $q$-expectation value of an observable, consistent
with the thermodynamic relations as we shall see in Eq.(\ref{uqz}) below. 

Working with the canonical ensemble, the probability distribution
can be obtained by extremizing the entropy $S_q$ under fixed internal energy
$U_q$ constraint and norm constraint ($\sum_i p_i=1$). The outcome of this
optimization procedure gives the result 
\begin{eqnarray}
p_i=
\frac{ \left[1-(1-q) \beta \epsilon_i
\right]^{\frac{1}{1-q}}}{Z_q} \; ,
\label{pq}
\end{eqnarray}
where $Z_q$ is the partition function given by
\begin{eqnarray}
Z_q&=&\sum_i \left[1-(1-q) \beta \epsilon_i
\right]^{\frac{1}{1-q}} \; .
\label{zq}
\end{eqnarray}

Using the generalized expression (\ref{logq}) of the logarithmic function and
Eqs.(\ref{uq}) and (\ref{zq}), it has been shown that \cite{curado}
\begin{equation}
U_q=-\frac{\partial}{\partial\beta}\;\ln_q Z_q\; .
\label{uqz}
\end{equation}
On the basis of the above relation, the entire
 mathematical structure of the connection between standard
statistical mechanics and thermodynamics is preserved by the generalization
of the Tsallis entropy, the definition of the internal energy and replacing
$\ln Z$ by $\ln_q Z_q$.

 Recently Abe \cite{abe} has
observed the connection between Tsallis entropy
and Jackson derivatives which can be expressed as

\begin{equation}
S_q=-\left. \djk \;\sum_i p_i^{x} \right|_{x=1}
\equiv -\sum_i\frac{p_i^q-p_i}{q-1}\; ,
\end{equation}
where
\begin{equation}
\djk f(x) = \frac{f(qx)-f(x)}{x \; (q-1)}\; ,
\label{djk}
\end{equation}
is the Jackson derivative \cite{jack}, which in the limit $q\rightarrow 1$,
becomes the ordinary differential.  
The above connection is not
just a coincidence but in fact it has been shown that the Jackson
derivative can be identified with the generators of fractal and multifractal
sets with discrete dilatation symmetries \cite{erz} and thus it is strictly
related to Tsallis statistics.

The Jackson derivative in Eq.(\ref{djk}) is intimately connected
with  $q$-deformed structures in $q$-oscillator theory, signified by the
$q$-basic number
\begin{equation}
[x]_q=\frac{q^x-1}{q-1} \; .
\end{equation}
It has been shown that the pseudo-additivity property of the
Tsallis entropy, displayed in Eq.(\ref{add}), is
also valid for the above $q$-basic number \cite{tsa}.

We now develop the $q$-symmetric theory of the Tsallis thermostatistics.
In $q$-deformed structures, when one constructs the theory which is invariant
under  $q\leftrightarrow \q1$, the Jackson derivative has to be generalized
 to the form
\begin{equation}
\dqq \;f(x) = \frac{f(qx)-f(\q1 x)}{x \; (q-\q1)}\; ,
\label{dqq}
\end{equation}
and correspondingly, it is possible to introduce the symmetric Tsallis entropy given by \cite{abe}
\begin{equation}
S_q^S=- \left. \dqq \;\sum_i p_i^{x}\right|_{x=1} =
- \sum_i \frac{p_i^q - p_i^\q1}{q-\q1}\; .
\end{equation}
The above expression for the symmetric Tsallis entropy satisfies 
a generalized pseudo-additivity property formally similar to the
symmetric $q$-basic number defined by
\begin{equation}
[x]_q^S=\frac{q^x-q^{-x}}{q-\q1} \; .
\end{equation}

It is easy to show that the $q$-symmetric Tsallis entropy can be written
in terms of the nonsymmetric Tsallis entropy in the compact form
\begin{equation}
S_q^S=c_1 \; S_q + c_2 \; S_\q1\; ,
\end{equation}
where $c_1$ and $c_2$ are two coefficients (always positive) which depend
only on $q$ and $\q1$,
\begin{equation}
c_1=\frac{q-1}{\qq1} \ \ \ , \ \ \ 
c_2=\frac{1-\q1}{\qq1} \ \ \ , \ \ \ c_1+c_2=1 \; .
 \end{equation}

We wish to stress that the above expression is a consequence of
 the connection between the $q$-symmetric and the nonsymmetric
  $q$-structures. In fact we observe 
that the Jackson
  symmetric derivative in Eq.(\ref{dqq}) can be expressed in terms
  of the nonsymmetric ones
\begin{equation}
\dqq=c_1 \; \djk +c_2 \; \partial_x^{(\q1)}\; .
\end{equation}
An analogous relation also holds for the $q$-basic number in
quantum groups
\begin{equation}
[x]^S_q = 
c_1 \; [x]_q + c_2 \;  [x]_\q1 \; .
\end{equation}
Accordingly the above relations provide us with a recipe to obtain the symmetric generalization from the nonsymmetric structures.

As in the case of Tsallis entropy, the equiprobability distribution
($p_i= 1/W\, , \;\forall{i}$)
can be derived by employing  a new symmetric definition
of the logarithmic function
\begin{equation}
S_q^S= \ln_q^S W= c_1 \ln_q W + c_2 \ln_\q1 W \; .
\end{equation}
The above result is very important for the correct construction of a
generalized symmetric thermostatistics. 
In fact this allows us to obtain the symmetric logarithm of the partition function (thermodynamic potential)
in terms of a linear combination of the Tsallis one
\begin{equation}
\ln_q^S Z_q=c_1 \; \ln_q Z_q +c_2 \; \ln_\q1 Z_\q1\; .
\end{equation}
Because Tsallis statistics satisfies the condition in Eq.(\ref{uqz}), 
it is also verified immediately in symmetric nonextensive statistics that 
\begin{equation}
U_q^S=-\frac{\partial}{\partial\beta}\;\ln_q^S Z_q \; ,
\end{equation}
if we choose the symmetric internal energy to be in the form
\begin{equation}
U_q^S=c_1 \; U_q +c_2 \; U_\q1 \equiv
\sum_i \epsilon_i\;\frac{(q-1)\, p_i^q -(\q1 -1)\, p_i^\q1}{q-\q1}\; .
\label{uqs}
\end{equation}

The definition in Eq.(\ref{uqs}) of the internal energy implies a generalized
symmetric $q$-expectation value of a physical  observable $A$
\begin{equation}
\langle A \rangle_q^S= c_1 \langle A \rangle_q+c_2 \langle A \rangle_\q1 
\equiv \sum_i A_i\;\frac{(q-1)\, p_i^q -(\q1-1)\, p_i^\q1}{q-\q1}\; ,
\end{equation}
where we note  that the second relation on the right hand side
is true only if  $A$ does not depend on $q$ (such as the case of
energy or  particle number).

The above results are very important because we have a new generalized
definition of the internal energy, which
together with the symmetric Tsallis
entropy, preserves all the thermodynamic relations (Legendre transformations). 
This follows directly from the   $q \leftrightarrow \q1$ invariance of the
$q$-deformed algebra,  thus offering  a closer connection between
nonextensive Tsallis statistics and $q$-deformed structures.

Following the standard procedure, the probability distribution
can be obtained by extremizing the entropy $S_q^S$ under fixed internal energy
$U_q^S$ constraint and the norm constraint ($\sum_i p_i=1$).
The result can be written as a linear combination 

\begin{equation}
p^S_i=c_1 \;
\frac{ \left[1-(1-q) \beta \epsilon_i
\right]^{\frac{1}{1-q}}}{Z_q}
+c_2 \;
\frac{ \left[1-(1-\q1) \beta \epsilon_i
\right]^{\frac{1}{1-\q1}}}{Z_\q1} \; .
\label{pqs}
\end{equation}
In the limit $q\rightarrow 1$, $p_i^S$ reduces to the standard
Maxwell-Boltzmann distribution. 
We note that the
extremization procedure only establishes that the solution for the
 distribution function is a linear combination of Tsallis distribution
 evaluated at $q$ and $\q1$.
 Eq.(\ref{pqs}) is a reasonable choice and follows  the prescription of the
 $q$-calculus.
In fact in the $q$-oscillator theory the statistical distribution function
 can be written as the same linear combination of the non-symmetric
 ones \cite{bie,song}
\begin{equation}
f_q^S=c_1 \; f_q+c_2 \; f_\q1 \; ,
\end{equation}
where $f_q$ and $f_\q1$ are the distribution functions in
non-symmetric $q$-boson oscillators 
\begin{equation}
f_q=\frac{1}{e^{\beta\omega}-q}\; .
\end{equation}

In Fig. 1, we show the plot of the normalized probability
function (\ref{pqs}) against $\beta \epsilon$ for different values of $q$.
Let us note that the above distribution has no cut-off as in
Tsallis's distribution for $q<1$ and the high energy tail of the distribution
is always enhanced compared  to the Maxwell-Boltzmann distribution since
$p_i^S$  has the following power law behavior at high energy, $
p_i^S  =  a \; E^{1/ (1-q)} + b \; E^{1/(1- \q1)}$.

In light of the above discussion, making a Legendre transform of the function
$\ln_q^S Z_q$ it is easy to verify the validity of the relation
\begin{equation}
S_q^S=\beta U_q^S+\ln_q^S Z_q \; ,
\end{equation}
which implies the standard thermodynamic relation
\begin{equation}
\frac{\partial S_q^S}{\partial U_q^S}=\frac{1}{T}
\end{equation}
and the $q$-deformed symmetric free energy is given by
\begin{equation}
F_q^S=-\frac{1}{\beta} \; \ln_q^S Z_q=U_q^S - T S_q^S \; .
\end{equation}
All the above equations reduce to the standard thermodynamic relations
in the limit $q \rightarrow 1$.

Finally, we note that Tsallis \cite{tsa3} recently introduced  a normalization
procedure for the $q$-expectation value of an observable in order to
remove some anomalies, such as non-additivity of the generalized
internal energy and non-invariance of the probability distribution
under the choice of origin of the energy spectrum.
In the framework of the new generalization, 
all the results of
the present investigation remain unaltered if we implement the normalization
according to
\begin{equation}
\widetilde{U}_q^S=  c_1 \widetilde{U}_q + c_2 \widetilde{U}_{\q1}\; ,
\end{equation}
where $\widetilde{U}_q$ is the normalized Tsallis internal energy
\begin{equation}
\widetilde{U}_q =
\frac{\sum_i \widetilde{p}_i^{\, q} \epsilon_i }{
\sum_i \widetilde{p}_i^{\, q} }\; ,
\end{equation}
and $\widetilde{p}_i$ is the modified Tsallis distribution in the
normalized $q$-expectation value given by \cite{tsa3}
\begin{eqnarray}
\widetilde{p}_i=\frac{ \left[1-(1-q) \beta (\epsilon_i-
\widetilde{U}_q)/\sum_j \widetilde{p}_j^{q}
\right]^{\frac{1}{1-q}}}{\widetilde{Z}_q}\; ,
\end{eqnarray}
with
\begin{eqnarray} 
\widetilde{Z}_q=\sum_i\left[1-(1-q) \beta (\epsilon_i- \widetilde{U}_q)
/{ \sum_j \widetilde{p}_j^{q}} \right]^{\frac{1}{1-q}} \; .
\end{eqnarray}

In summary, we have shown that it is possible to extend Tsallis
thermostatistics to the  $q \leftrightarrow \q1$ symmetric generalization
which preserves the Legendre transformation of standard thermodynamics. This
is achieved by introducing a $q$-symmetric expectation value which follows
directly from the extension of the nonsymmetric $q$-deformed theory to
the symmetric one. We thus establish a closer connection between nonextensive
Tsallis statistics and $q$-deformed structures. 

In conclusion, the relevance of the  $q \leftrightarrow \q1$ symmetry is
 well-known in $q$-oscillators from the mathematical structure as well as
 in applications \cite{mont,ang,ubri,pns}. Similarly we expect the symmetric
 Tsallis thermostatistics to be useful in many future investigations. 

\vspace{.2in}

\noindent
{\bf Acknowledgments}
\vspace{.1 in}

We are grateful to P. Quarati and C. Tsallis for reading this manuscript and for useful suggestions. One of us (A.L.) would like to thank the Physics Department of Southern Illinois University for warm hospitality where this work was done.

\begin{figure}[htb]
\mbox{\epsfig{file=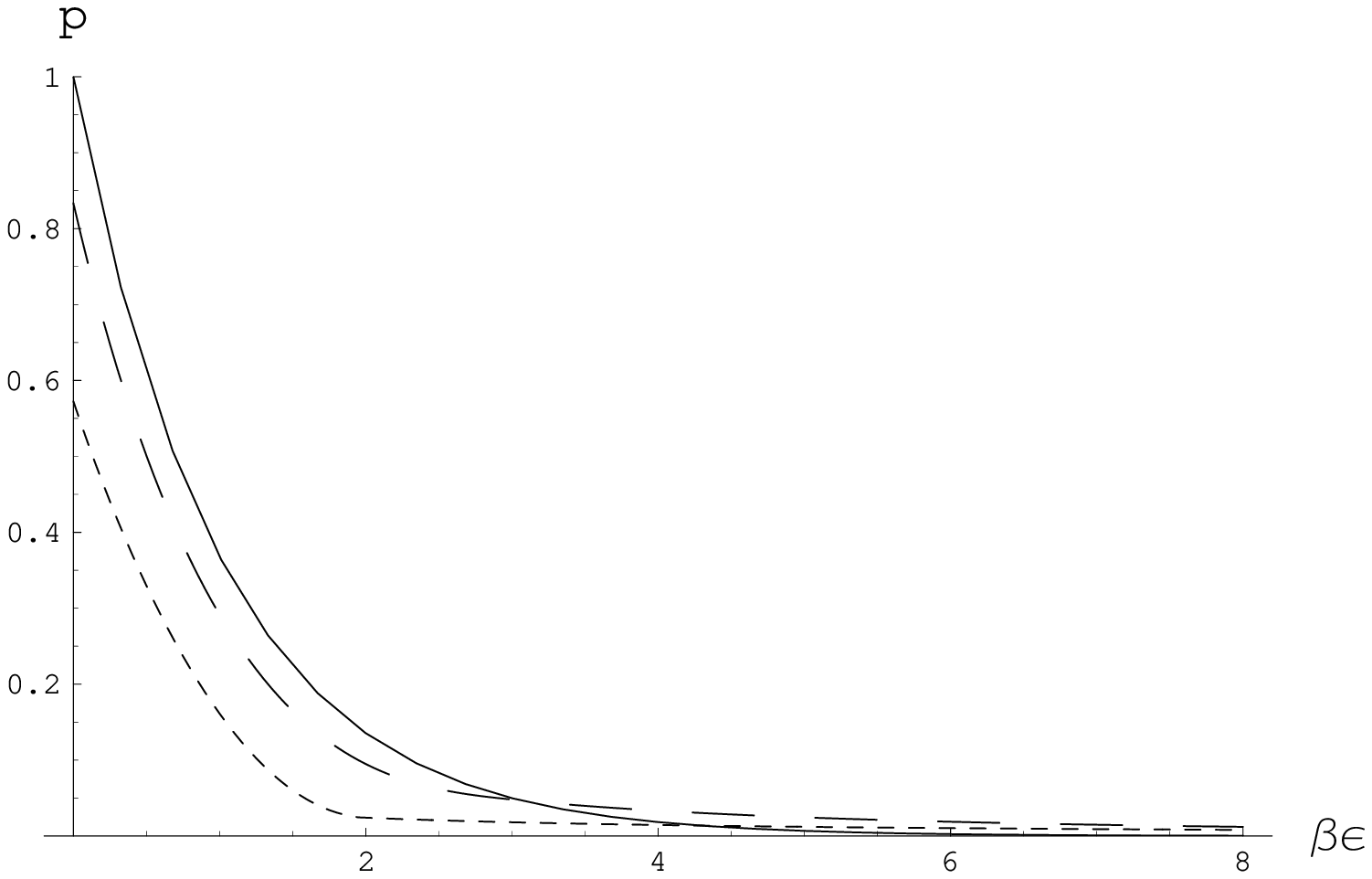,width=0.95\textwidth}}
\caption[]{Plot of the symmetric normalized probability
distribution, Eq.(\ref{pqs}), versus $\beta \epsilon$ for
$q=1.0$ (Maxwell-Boltzmann distribution), solid line; $q=1.5$
(equivalently $\q1=0.67$), long-dashed line and $q=2.0$
(equivalently $\q1=0.5$), short-dashed line.}
\end{figure}

\end{document}